\documentclass[12pt]{iopart}
\usepackage{graphicx,amssymb}
\begin{document}
\newcommand{\email}{\ead}
\newcommand{\affiliation}{\address}
\title{Natural clustering: the modularity approach}
\author{L. Angelini}
\email{angelini@ba.infn.it}
\author{D. Marinazzo}
\email{marinazzo@ba.infn.it}
\author{M. Pellicoro}
\email{pellicoro@ba.infn.it}
\author{S. Stramaglia}
\email{stramaglia@ba.infn.it}
 \affiliation{Dipartimento Interateneo di Fisica, Bari, Italy}
 \affiliation{TIRES, Center of Innovative Technologies for Image Detections and Processing, Bari, Italy}
 \affiliation{I.N.F.N., Sezione di Bari, Italy}
 \begin{abstract}
We show that modularity, a quantity introduced in the study of networked systems, can be
generalized and used in the clustering problem as an indicator for the quality of the solution.
The introduction of this measure arises very naturally in the case of clustering algorithms that
are rooted in Statistical Mechanics and use the analogy with a physical system.
 \end{abstract}
\maketitle

\section{Introduction\label{sec1}}
The problem of data clustering consists of grouping together items so that two points belonging to
the same group (cluster) are, in some sense, more similar than two that belong to different ones;
it has applications in several fields such as pattern recognition, bioinformatics, learning,
astrophysics and more (for a review see e.g. \cite{331504}). A data set composed by $N$ points is
represented either in terms of their coordinates (features) in a D-dimensional space or,
alternatively, by means of an $N\times N$ "distance matrix", whose elements measure the
dissimilarity of pairs of data points. Many clustering methods are now available. Some of them
make some assumption about the clusters' density distribution, so that so that clustering becomes
an optimization problem. Other approaches are called nonparametric as they do not impose a
determinate model to the data set.

 The clustering problem is inherently ill-posed, i.e. any data  set can be clustered in drastically
different ways, with no clear criterion to prefer one clustering over another. In particular in
the case of unsupervised approaches, a satisfactory clustering of data depends on the desired {\it
resolution} which determines the number of clusters and their size. However, the goal of data
clustering is to partition a sample according to \textit{natural} classes that are present in it.
For example in the case of gene clustering in microarray data one search to separate genes that
are involved in a particular biological processes from others which act as spectators. In the
features space the job consists of recognizing regions that are densely populated and are
separated by regions in which the density is lower.

The problem of \textit{natural} clustering becomes very important when analyzing the results of
clustering procedures. For example, if one applies different algorithms to the same data set what
is the criterion to choose among them the most efficient one? Let us consider yet another
situation: many algorithms perform hierarchical clustering, i.e. there is a parameter which
controls the resolution at which the data set is clustered; by varying the value of this parameter
the data set is grouped in a hierarchy of clusters ranging from the whole data set (one cluster)
to single items ($N$ clusters). Also in this situation the following question arises: what is the
best value of this parameter or, in other words, at what resolution should I look at the data to
find a scientific meaning in the classification?

 The problem of the most natural clustering solution has been explored in a recent paper
\cite{OtKeStSt2005}, where an answer is given in the frame of the Superparamagnetic clustering
(SC) method \cite{BlWiDo76,Do99}. The authors introduce an automated sequential procedure,
Sequential Superparamagnetic Clustering (SSC), in which SC is recursively applied: at each step
one well distinguished cluster is separated and in the next step SC is applied to the rest of the
sample. In this article we present an alternative procedure that emerges naturally in clustering
algorithms that are inspired by the analogy with a physical system, but can be applied in a
straightforward way also to other algorithms that perform hierarchical clustering.  The central
role in this approach is played by Modularity, a quantity which was first introduced in order to
detect community structure in networks; it is the object of next section. Then we briefly review
one of the clustering methods that are based on a Theoretical Physics approach and we show that
Modularity, if redefined starting from the coupling matrix rather than the adjacency matrix, is
able to select the most efficient cluster partition for a data set whose correct classification is
already known. Last section is dedicated to comments and conclusions.

\section{\label{sec:sec2}Networks and modularity} A network is the representation of a data system as
a set of nodes joined by edges that correspond to pairwise relations between nodes. Networks can
be used to describe many systems of scientific interest, such as electric power grids, road or
airline networks, the Internet and World Wide Web, social communities, biological or chemical
systems. The interest in the study of networks increased in the last decade together with the
availability of large databases, describing real world networks, and large computing power.
Recently many authors reviewed concepts and results in this discipline
\cite{AlBa02,Newman03,BoLaMoChHW06}, and we refer to them for further details.

 A typical feature of real networks is the appearance of tightly connected subgraphs with only
 sparser links between them. For example, in the case of social networks the existence of these
 subgroups could be related with the presence of important cultural differences between individuals.
 Many efforts have been dedicated to the problem of identifying communities in a network. One of
 the most effective method was introduced by Girvan and Newman \cite{Newman04,NeGi04}: their algorithm
 is able to produce a
 hierarchy of subdivisions in a network, from a single agglomerate to isolated nodes. They
 introduced also a quantity, the modularity Q, able to select which of the divisions is optimal.
 It is clear that, if a partition in a fixed number of subgroups of the network is requested, the
 best solution is obtained minimizing the number of edges connecting vertices belonging to
 different subgroups (or maximizing the number of vertices belonging to the same subgroup). But
 this is not a good recipe when the number of divisions is free, because this solution would
 correspond to no division at all. In this case a good division into communities is the one in
 which the number of edges between vertices belonging to the same group is significantly greater
 than the number expected from a random distribution of edges. This crucial concept is turned in
 mathematical quantity, the modularity.

 Let us consider a network composed by N vertices and an
 adjacency matrix $A$, where $A_{ij}$ is the number of edges ($0$ or $1$) between vertices $i$ and
 $j$. The number of edges incident with node $i$ is $k_i=\sum_j A_{ij}$ and the total number of
 edges is $m=\frac{1}{2}\,\sum_i k_i$. Suppose our network has been partitioned in $q$ groups; we
 can label each node $i$ with an index $\sigma_i\in\{1,2,\ldots,q\}$ denoting its group.
 Modularity is defined by
 \begin{equation}\label{mod}
    Q=\frac{1}{4m}\,\sum_{i\ne j}
    \left(A_{ij}-\frac{k_i\,k_j}{2m}\,\right)\delta(\sigma_i,\sigma_j)
 \end{equation}
The first term in the sum contributes to the number of edges falling within groups while the
second term is the expected number for the same quantity in case of a random distribution of
edges. Thus the problem of detecting communities in a network is turned into the problem of
maximizing the modularity. Many standard and dedicated optimization techniques have been
investigated; see \cite{Newman06} and references therein.

\section{\label{sec:sec3}Modularity and clustering algorithms}
Extending the use of modularity to the problem of data clustering, in particular when hierarchical
algorithms are used, requires the introduction of edges in the feature space where data are
represented. This comes out in a very natural way in the case of some algorithms that are rooted
in Theoretical Physics and that work through the analogy with a physical system. These algorithms
associate a physical quantity to each point in the features space, the entities beeing coupled by
a coupling constant that depends inversely by the distance. This interaction is able to drive the
system to an equilibrium condition where entities that are closer are characterized by a more
similar behavior. Clusters are then recognized by the measure of some variable identifying this
dynamical similarity. This is the case of an algorithm called Chaotic Map Clustering (CMC)
algorithm \cite{PRL00} to which some of the author of this paper contributed. We briefly review
the CMC algorithm.

We assign a real dynamical variable $x_i\in[­1,1]$ to each point of the data set and define
pair-interactions
\begin{equation}\label{couplings}
    J_{ij} = \exp \left\{-\frac{d_{ij}}{2 \alpha^2 \langle d_{ij}\rangle^2}\right\},
\end{equation}
where $d_{ij}$ is a suitable measure of distance between points i and j in our D-dimensional
space, $\langle d_{ij}\rangle^2$ is its average over the sample and $\alpha$ is a parameter. The
time evolution of the system is given by:
\begin{equation}\label{evolution}
    x_i (t + 1) = \frac{1}{C_i}\sum_{j\ne i} J_{ij}\; f(x_j(t)),
\end{equation}
where $C_i = \sum_{j\ne i} J_{ij}$, and we choose the map $f(x) = 1 - 2x^2$. Due to the choice of
the function $f$, equations (\ref{evolution}) represent the dynamical evolution of chaotic maps
$x_i$ coupled through pair interactions $J_{ij}$. The lattice architecture is fully specified by
fixing the value of $\alpha$ as the average distance of $k$-nearest neighbors pairs of points in
the whole system. For the sake of computational economy, we consider only interactions of each map
with a limited number of maps whose distance is less than $3\alpha$, and set all other $J_{ij}$ to
zero. Starting from a random initial configuration of $x$, equations (\ref{evolution}) are
iterated until the system attains its stationary regime, corresponding to a macroscopic attractor
which is independent of the initial conditions. To study the correlation properties of the system,
we consider the mutual information $I_{ij}$, between pairs of variables whose definition
\cite{WIG} is as follows. If the state of element $i$ is $x_i (t) > 0$ then it is assigned a value
$1$, otherwise it is assigned $0$: this generates a sequence of bits, in a certain time interval,
which allows the calculation of the Shannon entropy $H_i$ for the $i$­th map. In a similar way the
joint entropy $H_{ij}$ is calculated for each pair of maps and finally the mutual information is
given by $I_{ij} = H_i +H_j -H_{ij}$. The mutual information is a good measure of correlations and
it is practically precision independent, due to the coarse graining of the dynamics. If maps $i$
and $j$ evolve independently then $I_{ij} = 0$; if the two maps are exactly synchronized then the
mutual information achieves its maximum value, $\ln 2$. The algorithm identifies clusters with the
linked components of the graph obtained by drawing a link between all the pairs of maps whose
mutual information exceeds a threshold $\theta$: $\theta$ controls the resolution at which data
are clustered. Hierarchical clustering is obtained repeating this procedure for an increasing
sequence of $\theta$-values: each clustering level is extracted as a partition of data with a
finite stability region in the $\theta$ parameter. The algorithm is very fast, also for huge data
set requiring much computer memory, due to limit in the couplings that renders the interaction
local. The two task, coupled maps evolution and hierarchical cluster tree building can be
separated for computational needs. With regard to the choice of $\alpha$ we note that for small
values the lattice is very fragmented, while for large values the dependencies of the coupling
(\ref{couplings}) from the distance becomes flat and the algorithm is unable to reveal
substructures in the data set. Analyzing data sets for which the expected classification is known,
we got good results for $\alpha$ between $0.10$ and $0.30$, the algorithm being quite insensitive
in this range.

In precedent applications of CMC algorithm we considered as optimal partition of the given data
set the most stable solution, i.e. the one ranging on the widest $\theta$ interval. However the
capacity of modularity to reveal the underlying structure in networks suggests to extend use of
this quantity also for data mining. There are two ways to introduce modularity in our algorithm,
depending on the definition of edge between to elements of the data set. The first way is to
replace the adjacency matrix element $A_{ij}$, giving the link between site $i$ and site $j$, with
the value of the coupling $J_{ij}$ between the two maps $x_i$ and $x_j$. The matrix $J$ is also a
sparse matrix, but differs from matrix $A$ because its elements have real values. One can think of
the data set in the feature space as a network with weighted edges. Another way consists of the
use of the mutual information matrix $I$ measuring the synchronization between the maps. This
second method is more algorithm-depending, but has the practical advantage that computer memory
can be freed deallocating the coupling matrix in the final stage. We applied both methods on many
data sets that are currently used to test clustering algorithms. In every case we found that the
maximum of modularity corresponded to the highest value of efficiency (percentage of correctly
classified points). Modularity was calculated following both methods, coupling and mutual
information, and we did not find significative differences between them.

\begin{figure}[ht]
\begin{center}
\includegraphics[width=12.5cm]{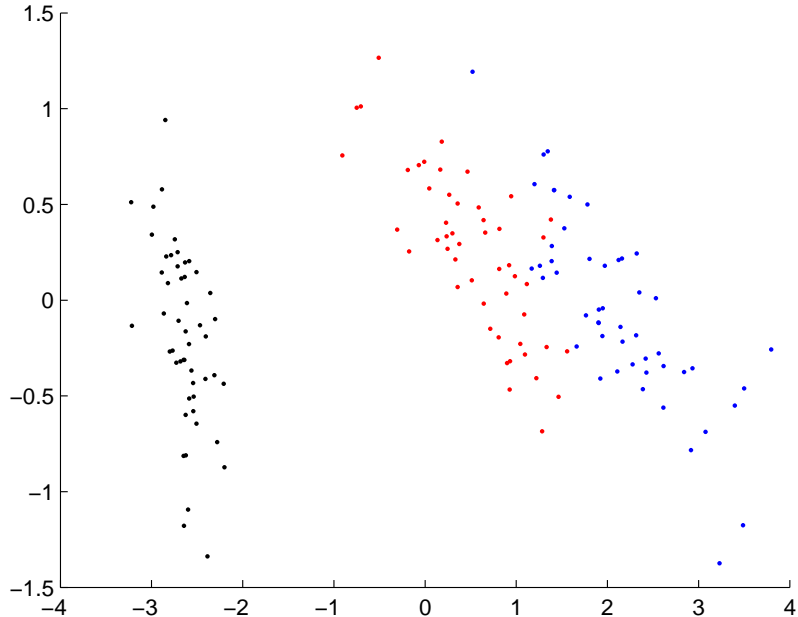}
\end{center}
\caption{The Iris data set projection in the first two principal components plane} \label{iris}
\end{figure}

\begin{figure}[ht]
\begin{center}
\includegraphics[width=12.5cm]{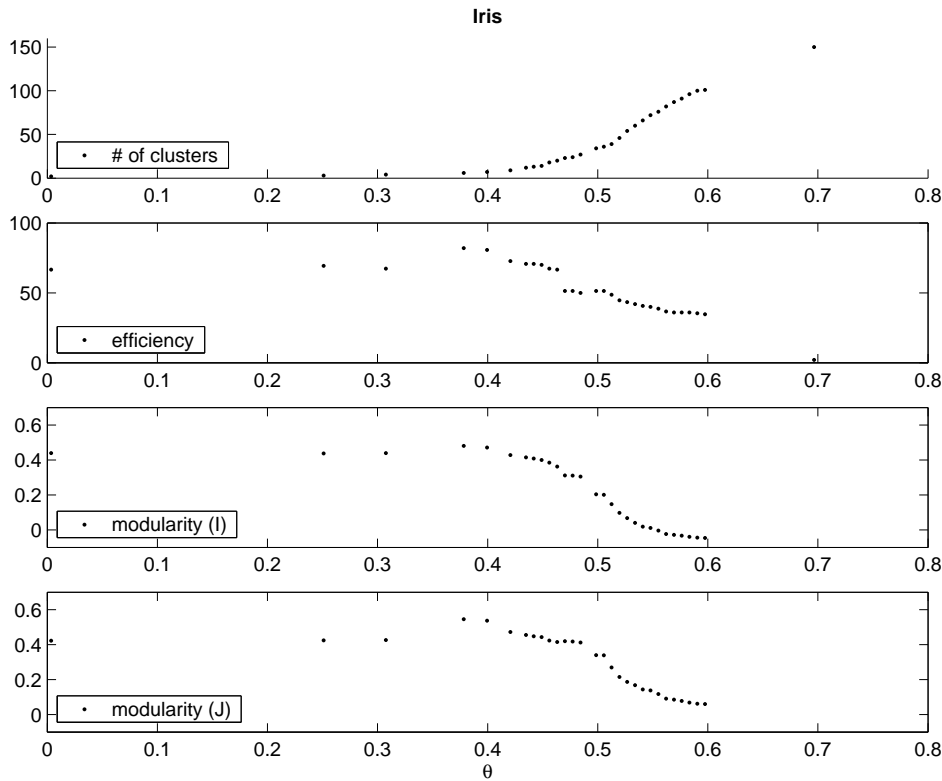}
\end{center}
\caption{Efficiency and modularity for the Iris data sample} \label{irismod}
\end{figure}

 As an example we report here results from the application of CMC algorithm (with Euclidean distance)
to the famous IRIS data sample of Anderson \cite{anderson}. This four dimensional data set, which
is represented in the two first principal component plane in fig.(\ref{iris}), has often been used
as a standard for testing clustering algorithms: it consists of three clusters (Virginica,
Versicolor and Setosa) and there are $50$ objects in ${\mathbf R}^4$ per cluster.
Two clusters (Virginica, Versicolor) are very overlapping.
 Fig. (\ref{irismod}) shows, as a function of $\theta$, the following
quantities (from top to bottom): the number of clusters, the efficiency, the modularity calculated
using the couplings matrix $J$, the modularity calculated using the mutual information matrix $I$.
(Points are not equidistant because we add a new point only when the number of clusters changes.)
 We point out the similarity between the behavior of the efficiency, that measures the goodness
of the classification, and each of the two modularity's, in particular with respect to their
maxima position. For $\theta=0.423$ the two modularity's get their maximum value together with the
maximum value of the efficiency, $82\%$.

\begin{figure}[ht]
\begin{center}
\includegraphics[width=12.5cm]{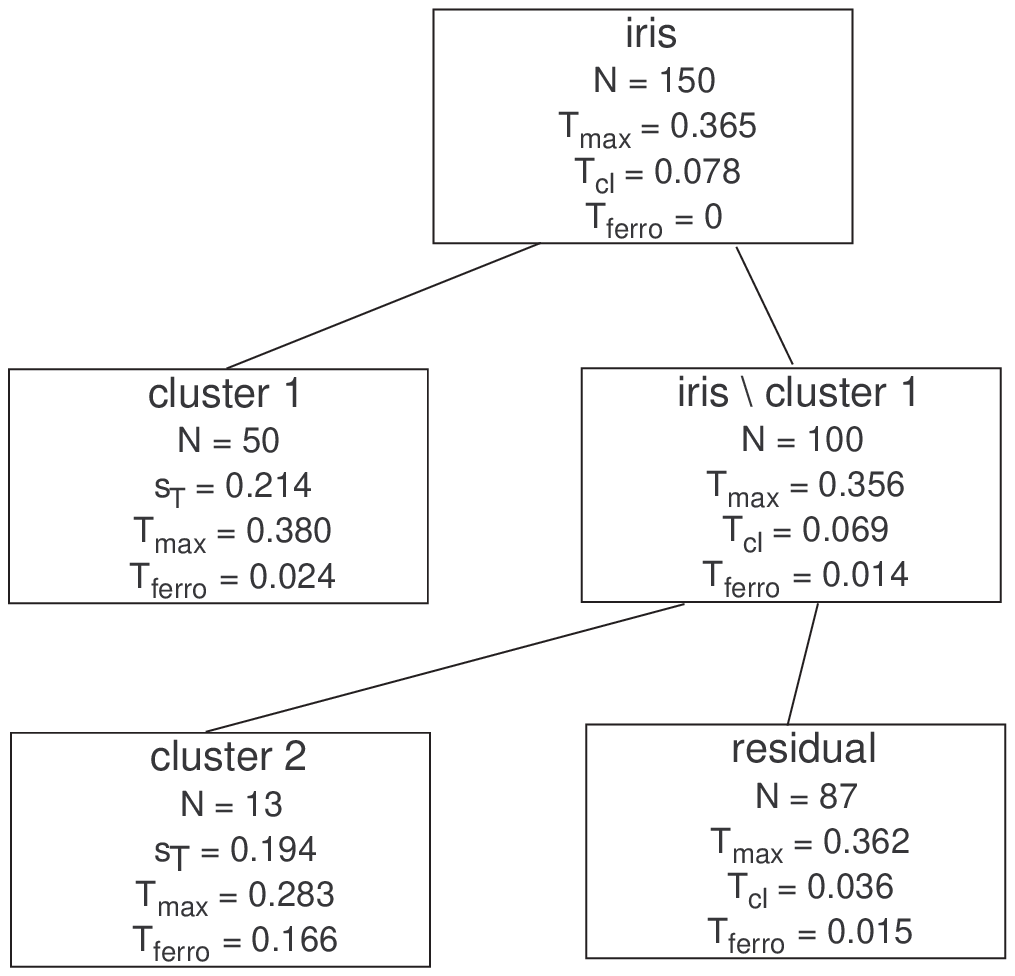}
\end{center}
\caption{Cluster structure resulting from the application of SSC algorithm \cite{OtKeStSt2005} to
the Iris data set.} \label{iris_stoop}
\end{figure}

We also applied to the same data set the SSC method \cite{OtKeStSt2005}. Results from this test
are shown in figure \ref{iris_stoop}; to explain them we shortly describe this algorithm.
SSC is an extension of Superparamagnetic Clustering of Domany and coworkers \cite{BlWiDo76}, in
which every point $i$ of the data set to be analyzed is associated with a Potts spin variable
$s_i\to\{1,2,...,q\}$, where $q$ is of the order of tenths. Each spin is coupled to its $k$
nearest neighbors with a coupling which expression is similar to \ref{couplings}. The system
evolves to equilibrium configurations trough a canonical Monte Carlo procedure at temperature T
and two points are considered to belong to the same cluster if their pair correlation function is
larger than a threshold value. Depending from the temperature $T$, the system can be found in
three phases: a ferromagnetic phase from $T=0$ to $T=T_{ferro}$, a paramagnetic phase for $T$
greater than a value $T=T_{max}$ and a superparamagnetic phase at intermediate $T$. In the
superparamagnetic phase only closest spins are strongly correlated, so that this phase is
characterized by the formation of clusters. Two spins $i$ and $j$ are attributed to the same
cluster if the pair correlation $G_{ij}=\langle \delta_{s_i,s_j}\rangle$ exceeds a fixed threshold
$\Theta$. The number of these clusters increases with $T$ from $T_{ferro}$ to $T_{max}$, resulting
in a classification hierarchy, so that temperature acts as the parameter that controls the
resolution  at which data are clustered. This method has some difficulty to deal with data sets
that contain clusters with different densities, because these clusters do not appear at the same
resolution (temperature) level. Sequential Supermagnetic Clustering tries to solve this problem
through a sequential procedure in which at each step the most stable cluster is extracted from the
data set. Then the most stable cluster and the residual set are clustered separately. This
procedure is applied repeatedly resulting in a binary tree. Cluster stability here is temperature
stability, i.e.
\begin{equation}\label{stability}
    s_T=\frac{T_{cl}}{T_{max}}
\end{equation}
as stability parameter is used, where $T_{cl}$ is the temperature range over which the cluster
emerges. At each level of the procedure the branching process is stopped if clustering produces
subcomponents none of which has stability greater than a threshold value $s_{\Theta}$.
$T_{ferro}$, the temperature at which the cluster separates, should not be very small for a
cluster to be considered as a natural cluster. This is true also for the final branches: in this
case if $T_{ferro}$ takes a vanishing value the points of the cluster are considered as noise.

When applying SSC algorithm to the Iris data set we used for the number of Potts states the value
$q=20$, for the number of nearest neighbors the value $k=10$ and for the stability threshold the
value $s_{\Theta}=0.1$ (but results are stable in the range $0.1 \div 0.194$). As we see in figure
\ref{iris_stoop}, at the first step the method succeeds in separating the Setosa cluster from the
Virginica and Versicolor points, but it is unable to separate these two species efficiently.
Lowering the stability threshold under 0.099, the $87$-points cluster splits in two clusters of 81
and 6 points, so that results get worse. It seems that SSC solves very smartly the problem of
identifying homogeneous classes, but it meets difficulties when applied to recognize classes which
present inhomogeneities inside, even though composed of elements that are close in the features
space.

\begin{figure}[ht]
\begin{center}
\includegraphics[width=12.5cm]{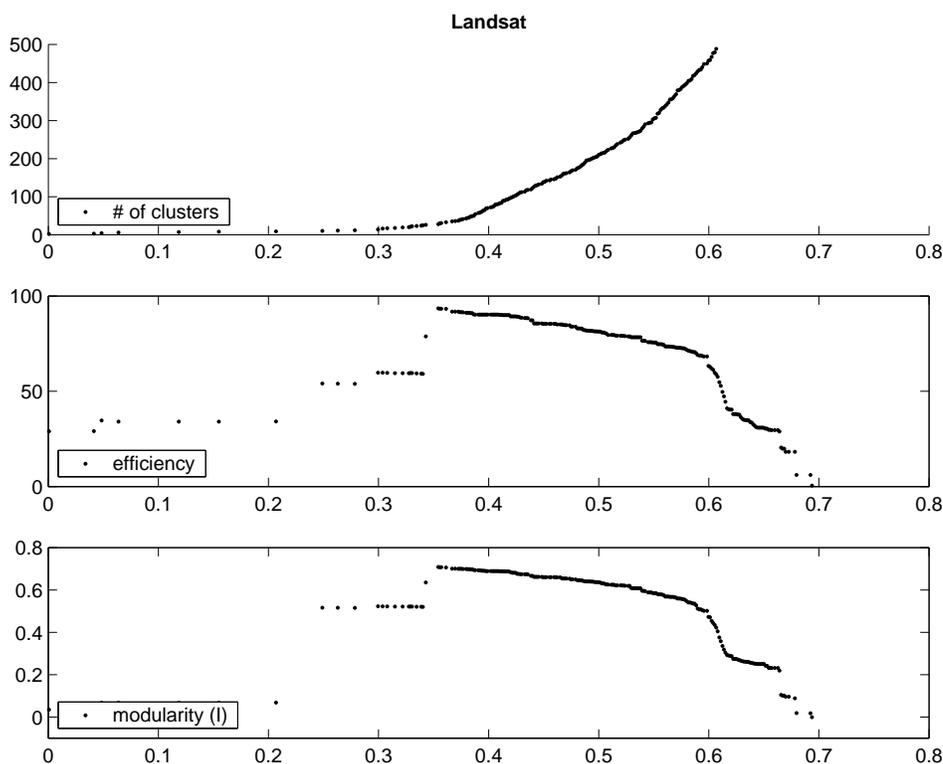}
\end{center}
\caption{Efficiency and modularity for the LANDSAT data sample} \label{landscapemod}
\end{figure}

Now we go back to the modularity approach and present another example given by the data set used
in \cite{PRL00}, that is also characterized by clusters of different size and density.
\begin{figure}[ht]
\begin{center}
\includegraphics[width=12.5cm]{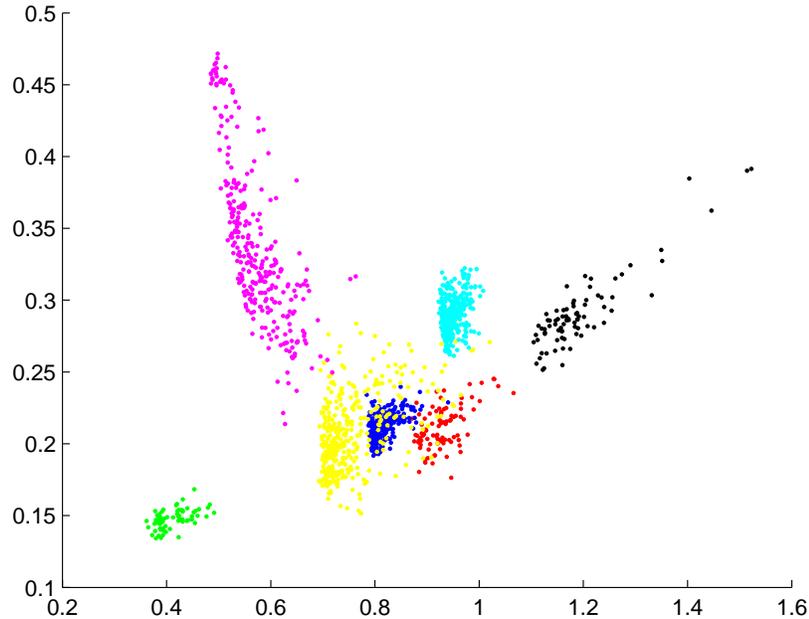}
\end{center}
\caption{The LANDSAT TM image data set projection in the first two principal components plane.
Colors distinguish among different landcover classes.} \label{landsat}
\end{figure}
It is a real data-set extracted from a LANDSAT Thematic Mapper (TM) image of an area in the
Southern of Italy consisting of $1489$ pixels each of which is represented by six spectral bands.
The ground truth was determined by means of visual interpretation of areal photos followed by site
visits. The area study includes seven landcover classes: (A) {\it Coniferous reafforestation},
$69$ points; (B) {\it Bare soil}, $85$ points; (C) {\it Urban areas}, $91$ points; (D) {\it
Vineyards}, $300$ points; (E) {\it Cropland}, $316$ points; (F) {\it Pasture}, $265$ points; (G)
{\it Olives groves}, $363$ points. As reported in \cite{PRL00} our algorithm succeeded in
resolving the data-structure identifying seven cluster that where stable on a large range of
$\theta$; $96.6\%$ of data was classified; the purity of the classification (percentage of
correctly classified points) is $96.2\%$ so that the efficiency was $92.9\%$. We remember also
that we searched for the best classification requiring that it would be the one surviving for  the
largest range in $\theta$. We found that both the six clusters partition and the seven clusters
one appeared stable: prior knowledge was needed, in this case, to select the correct partition of
the data set. We reanalyzed this data set according to the criterion of maximum modularity. The
results from these analysis are essentially the same as those reported previously: as we see in
fig (\ref{landscapemod}) modularity reached its maximum value of $0.71$ at $\theta=0.35$ in
presence of $7$ big clusters and $19$ small clusters of few points, the efficiency being $93.49$.
This slight improvement in itself does not justify the introduction of the modularity criterion;
however we note that, for this data set, the previous analysis had such a good performance that it
was not possible to improve much further. Finally, we emphasize that the problem of choose
choosing between the two stable solutions found an optimal solution without an external
supervision.

\section{\label{sec:sec4}Conclusions}
 We have shown that the problem of finding the optimal classification in hierarchical clustering
 can be turned into the problem of finding communities in a \textit{weighted network}.
 Results reported in this article  are obtained using the CMC algorithm, but it should be clear that the modularity
 method can be applied to any hierarchical clustering algorithm. For example it can be applied also to the
 Super-paramagnetic clustering (SPC) approach. Here the role of $\theta$ is played by temperature, but
 couplings between the elements of the data set have the same definition.
 We stress two important characteristic of the method here proposed. The first one is that no new parameter is introduced.
 The second feature is that it reduces considerably the computational time with respect to the alternative
method described in \cite{OtKeStSt2005}, where one needs many Monte Carlo runs to find just one
cluster. We remember also that, in that method, once one cluster is identified, it is eliminated
from the sample and this could result in the loss of the collective aspect of the data
distribution.

We are aware that the problem of finding the most natural clusters has not a unique solution, but
it seems to us that the modularity maximum method corresponds to an intuitive idea of optimal
classification as the one in which connectivity within groups is as far as possible the one
obtained from a random distribution.
 \section*{References}
 \bibliographystyle{unsrt}
\bibliography{mybibfile}

\end{document}